\documentstyle[12pt,psfig,epsf]{article}

\newcommand{\rf}[1]{(\ref{#1})}
\newcommand{\bea}{\begin{eqnarray}}
\newcommand{\eea}{\end{eqnarray}}

\newcommand{\g}{\gamma}
\renewcommand{\l}{\lambda}

\renewcommand{\a}{\alpha}

\newcommand{\ep}{\varepsilon}

\newcommand{\del}{\delta}
\newcommand{\sg}{\sigma}

\newcommand{\oh}{\frac{1}{2}}

\newcommand{\ra}{\right\rangle}
\newcommand{\la}{\left\langle}
\newcommand{\hstackrel}[2]{ 
    \stackrel{\hbox{\scriptsize #1}}{\hbox{\scriptsize #2}} }

\newcommand{\cT}{{\cal T}}
\def\void{}
\def\labelmark{}

\newenvironment{formula}[1]{\def\labelname{#1}
\ifx\void\labelname\def\junk{\begin{displaymath}}
\else\def\junk{\begin{equation}\label{\labelname}}\fi\junk}%
{\ifx\void\labelname\def\junk{\end{displaymath}}
\else\def\junk{\end{equation}}\fi\junk\labelmark\def\labelname{}}

{\ifx\void\labelname\def\junk{\end{array}\end{displaymath}}
\else\def\junk{\end{array}\right.\end{equation}}
\fi\junk\labelmark\def\labelname{}\def\junk{}
}

\newcommand{\beq}{\begin{formula}}
\newcommand{\eeq}{\end{formula}}
\newcommand{\beqv}{\begin{formula}{}}

\begin{document}
\topmargin 0pt
\oddsidemargin 5mm
\headheight 0pt
\headsep 0pt
\topskip 9mm

\hfill    NBI-HE-96-64

\hfill    TIT/HEP--352

\hfill    November 1996

\begin{center}
\vspace{24pt}
{ \large \bf Quantum geometry of topological gravity}

\vspace{24pt}

{\sl J. Ambj\o rn}${\,}^1$,  
{\sl K. N. Anagnostopoulos}${\,}^1$, 
{\sl T. Ichihara}${\,}^2$, 
{\sl L. Jensen}${\,}^1$, \\
{\sl N. Kawamoto}${\,}^3$, 
{\sl Y. Watabiki}${\,}^2$ and {\sl K. Yotsuji}${\,}^3$

\vspace{36pt}

${}^1$ The Niels Bohr Institute, University of Copenhagen, \\
Blegdamsvej 17, DK-2100 Copenhagen \O , Denmark

\vspace{24pt}

${}^2$ Department of Physics, Tokyo Institute of Technology,\\ 
{\O}-okayama, Meguro, Tokyo 152, Japan

\vspace{24pt}

${}^3$ Department of Physics, Hokkaido University,\\
Sapporo, Japan
\end{center}
\vspace{24pt}

\vfill

\begin{center}
{\bf Abstract}
\end{center}

\vspace{12pt}

\noindent
We study a $c=-2$ conformal field theory coupled to two-dimensional
quantum gravity by means of dynamical triangulations.  We define the
geodesic distance $r$ on the triangulated surface with $N$ triangles,
and show that ${\rm dim}[ \, r^{d_H} \, ] = {\rm dim}[ \, N \, ]$,
where the fractal dimension $d_H = 3.58\pm0.04$.  This result lends
support to the conjecture $d_H = -2\alpha_1/\alpha_{-1}$, where
$\alpha_{-n}$ is the gravitational dressing exponent of a spin-less
primary field of conformal weight $(n+1,n+1)$, and it disfavors the
alternative prediction $d_H = -2/\gamma_{\rm str}$.  On the other
hand, we find ${\rm dim}[ \, l \, ] = {\rm dim}[ \, r^2 \, ]$ with
good accuracy, where $l$ is the length of one of the boundaries of a
circle with (geodesic) radius $r$, i.e. the length $l$ has an
anomalous dimension relative to the area of the surface.  It is
further shown that the spectral dimension $d_s = 1.980\pm0.014$ for
the ensemble of (triangulated) manifolds used.  The results are
derived using finite size scaling and a very efficient recursive
sampling technique known previously to work well for $c=-2$.

\vfill

\newpage

\section{Introduction}

Liouville theory and matrix models have been successful in explaining 
a number of features of conformal field theories
coupled to two-dimensional quantum gravity. 
However, our primary interest in a theory of quantum gravity concerns 
geometry: which concepts of geometry survive the quantum average,
and how is the geometry changed by this average. 
These questions have not been clarified by Liouville field theory 
or matrix model techniques.
In the last couple of years significant progress has been made 
in this direction, starting with the introduction of the so-called 
transfer matrix \cite{transfer}.
It was shown that a reparametrization invariant formulation of 
the two-point function in quantum gravity has a simple geometric 
interpretation \cite{aw} and that a generalization 
of the two-point function to include matter fields allows 
a geometric interpretation of the {\it KPZ}-exponents \cite{aamt}, 
an interpretation first conjectured in \cite{ajw}. 
In addition it was realized that finite size scaling analysis 
of the two-point functions were very efficient tools 
for extracting critical exponents \cite{suracuse,ajw}. 
In this article we will take advantage of this new technology 
and combine it with the efficient recursive sampling algorithm 
developed earlier for $c=-2$ conformal field theory coupled to 
two-dimensional quantum gravity used in \cite{kketal} 
where the fractal nature of quantum gravity in two dimensions 
was first numerically confirmed. 

\section{The model}

The $c=-2$ model coupled to quantum gravity corresponds 
to a non-unitary $(1,2)$ conformal field theory coupled to quantum gravity, 
known as a topological quantum gravity. 
There exists an explicit realization of the model 
within the framework of dynamical triangulations. 
In this framework the partition function for $c$ Gaussian fields 
coupled to two-dimensional quantum gravity is 
\beq{*1}
Z_N  =  \sum_{\cT_N}  \left(\det C_{\cT_N}\right)^{-c/2}\, .
\eeq
In \rf{*1} the summation is over all triangulations $\cT_N$ with fixed topology
(which we will always assume to be spherical in this paper) 
built from $N$ triangles, 
and $C_{\cT_N}$ is the so-called adjacency matrix of the graph 
corresponding to the triangulation $\cT_N$. 
Notice that the triangulations of the spherical surface 
are in one-to-one correspondence with 
the $\phi^3$ connected planar graphs with no external legs, 
and thus it is possible to generate any triangulation $\cT_N$ 
if we generate and connect $\phi^3$ trees and rainbow diagrams 
with the correct weight. 
If $c=-2$, 
the weight for generating $\phi^3$ trees and rainbow diagrams 
is $1$, 
i.e. Eq.\ \rf{*1} can be written \cite{kkm} 
\beq{*2}
Z_N  =  \frac{1}{N+2} 
\Bigl( \sum_{ \hstackrel{tree diagrams}{with $N+2$ legs} } 1 \, \Bigr) 
\Bigl( \sum_{ \hstackrel{rainbow diagrams}{with $(N+2)/2$ lines} } 1 \, \Bigr) 
\, ,
~~~~~\hbox{(for $c=-2$)}, 
\eeq
where the first summation is over all rooted $\phi^3$ tree diagrams 
with $N+2$ external legs 
and the second summation is over all rainbow diagrams 
with $(N+2)/2$ lines. 
$1/(N+2)$ is the symmetry factor which comes from 
connecting the tree diagram and the rainbow diagram. 
Using a recursive algorithm for generating $\phi^3$ trees 
and the rainbow diagrams in \rf{*2}, 
it is possible to create a large number of independent triangulations
$\cT_N$ with the weight $\det C_{\cT_N}$. 
We refer to \cite{kketal} and \cite{us} for details. 
Supplementary to the first studied in \cite{kketal} 
we will in this work study the finite size aspects of the observables
associated with $Z_N$, 
the main motivation being that finite size scaling by far 
is the most reliable method for extracting continuum physics 
in critical systems and we now understand that this is true
also for these systems coupled to quantum gravity.

One very important point in the above setup is that 
we have the concept of distance, 
even if we usually associate $c=-2$ model with a topological gravity. 
To a triangulation $\cT_N$ 
we can unambiguously associate a piecewise linear manifold with 
a metric dictated by the length assignment $\varepsilon$ to each link. 
{}From a practical point of view 
we use instead a graph-theoretical distance between vertices,
links or triangles. 
In the limit of very large triangulations we expect that 
the different distances when used in ensemble averages
will be proportional to each other. 
To be specific we will in the following operate with 
a ``link distance'' and a ``triangle distance''.  
The link distance between two vertices is defined as the
shortest link-path between the two vertices, 
while the triangle distance between two triangles is defined as the 
shortest path along neighboring triangles between the two triangles. 
In this way the triangle distance becomes the link distance 
in the dual $\phi^3$ graph.

In the following we will report on the measurement of two quantities
related to the fractal structure of quantum space-time: 
The total length $\la l \ra$ (and the higher moments $\la l^n \ra$) 
of boundaries of spherical balls of (geodesic) distance $r$, 
and the measurement of so-called spectral quantities, 
originating from the study of random walks on the manifolds.
 
Let us define the observables on the triangulation $\cT_N$ more precisely. 
We consider a spherical ball of radius $r$ and its shell 
for a given triangulation $\cT_N$. 
The spherical ball consists of all vertices with link distance $r' \leq r$ 
and the spherical shell consists of all vertices with link distance $r$, 
where the distance is measured from a given vertex $v_0$ 
which is considered as the center of the spherical ball. 
In the same way 
we can define the spherical shell in terms of triangle distance. 
We will use both definitions in the following. 
The spherical shell will in general consists of 
a number of connected components. 
If we take the average over all positions of $v_0$ 
and all triangulations $\cT_N$, 
we get a distribution $\rho_N(l,r)$ of the length $l$ 
(measured in link units) of the connected components of 
the spherical shells of radius $r$, 
i.e.  
\beq{*3} 
\la l^n \ra_{r,N}  \equiv  \sum_{l=1}^\infty l^n \rho_N(l,r)\, .  
\eeq 
In particular we introduce the special notation
$n_N(r) = \la l(r) \ra_N$ and expect the fractal dimension to be
related to $n_N(r)$ by
\beq{*4} 
n_N(r) \sim r^{d_H-1},~~~~~~~1 \ll r \ll N^{1/d_H}\,.  
\eeq 
According to general scaling arguments \cite{aw,suracuse,ajw}
we expect the following behaviour for $n_N(r)$: 
\beq{*5} 
n_N (r) \sim N^{1-1/d_H} F_1(x),~~~~~x = \frac{r}{N^{1/d_H}}\, , 
\eeq 
and we expect $F_1(x)$ to fall off rapidly when $x \gg 1$.

The spectral properties are derived from the study of random 
walks on the triangulated surfaces via the diffusion equation
\beq{*6}
\phi(v,t+1) = \frac{1}{n_v} \sum_{(vv')} \phi(v',t)\, ,
\eeq
where $t$ is the diffusion time and the summation is over the $n_v$
neighboring vertices\footnote{Also in this case we have the
possibility to formulate a diffusion in terms of triangles and triangle
distances, rather than vertices and link distances.  However, 
the finite size effects are larger for triangles and we will in the
following only use link distances when we discuss diffusion.} $v'$ to $v$. 
We here consider the initial condition of $\phi(v,t)$, 
\beq{*7}
\phi(v,0) = \frac{1}{n_v} \del_{v,v_0}\, ,
\eeq
where $v_0$ is a fixed vertex. 
Eqs.\ \rf{*6} and \rf{*7} are 
discretizations of the diffusion equation on a continuum manifold. 
We will be interested in observables associated with the diffusion process 
obtained by averaging over the chosen vertex $v_0$ 
as well as different triangulations $\cT_N$.  
In the following discussion 
we will always assume that this average has been performed. 
Let us denote the probability of diffusion out to a link distance $r$ 
in time $t$ by $k_N(r,t)$.
By definition we have 
\beq{*8}
\sum_{r=0}^\infty n_N(r) \, k_N(r,t) = 1\, .
\eeq
{}From scaling arguments \cite{ksw,ajw,watabiki2}
we expect the following scaling
\beq{*9}
k_N(r,t) = \frac{1}{N}\; p(x,y),~~~~x = \frac{r}{N^{1/d_H}},~~y = \frac{t}{N^\l} \, ,
\eeq
where the new exponent $\l$ is defined such that $y$ will be finite 
in the scaling limit. The {\it spectral dimension} is defined 
from the return probability by
\beq{*10}
k_N(0,t) \sim  \frac{N^{\l d_s/2-1}}{t^{d_s/2}} \, ,
~~~~~{\rm for}~~~t \sim 0 \, ,
\eeq
while the average geodesic distance travelled by diffusion at time $t$ is 
\beq{*11}
\la r(t) \ra_N \equiv \sum_{r=0}^\infty  r \; n_N (r) k_N(r,t) 
\sim  N^{1/d_H - \l \sg} t^\sg \, ,
~~~~~{\rm for}~~~t \sim 0\, .
\eeq
If a kind of \lq\lq smooth'' fractal is expected here again, 
$k_N(0,t)$ and $\la r(t) \ra_N$ exist 
and are different from zero in the limit $N \to \infty$. 
This implies the scaling relations 
\beq{*smoothcond2}
 d_s = \frac{2}{\l} \, ,~~~~~ \sg = \frac{1}{\l d_H} \, .
\eeq  

\section{Numerical results}

\subsection{The simulations}

The simulations are performed by generating a number of statistically
independent configurations using the algorithm mentioned in the
introduction (see \cite{kketal} and \cite{us} for details). We use the
high quality random number generator RANLUX \cite{ml,fj} whose
excellent statistical properties are due to its close relation to the
Kolmogorov K-system originally proposed by Savvidy
et.al.\cite{ss,ass} in 1986\footnote{The history of the seminal paper
by Savvidy et.al. is interesting: The paper was rejected by 4 computer
journals, including Comput. Phys. Commun. where M. L\"uscher finally
published a related paper and F. James the FORTRAN code of RANLUX in
1994. These ideas were communicated to F. James by G.K. Savvidy during
his stay at CERN in 1987.}. We report results on system sizes ranging
from $2000${--}$256000$ triangles. The number of configurations
obtained depends on the lattice size and on the observable that we
measure. We choose $20$ random vertices on each configuration in order
to perform correlation function measurements. 
We need to collect more statistics to test Eq.\ \rf{*5}, 
where we have between $4.2\times 10^6$ and $1.6\times 10^6$ 
configurations. 
For the $128K$ and $256K$ lattices we have 
$6\times 10^5$ and $2\times 10^5$ configurations respectively. 
In order to measure the moments $\la l^n \ra_{r,N}$ 
and their scaling properties 
we need a factor of $10^2$ less configurations: 
We have approximately $50000$ configurations for each lattice size. 
Unfortunately, the computer effort for making the measurements is 
comparable to the one needed to test Eq.\ \rf{*5} with enough accuracy. 
For the diffusion equation we collect $2500$ configurations 
on which we perform $5$ measurements. 
For the three largest lattices we have $620$, $600$ and $400$ 
configurations respectively.

\subsection{The fractal dimension}
\begin{figure}
\centerline{\hbox{\psfig{figure=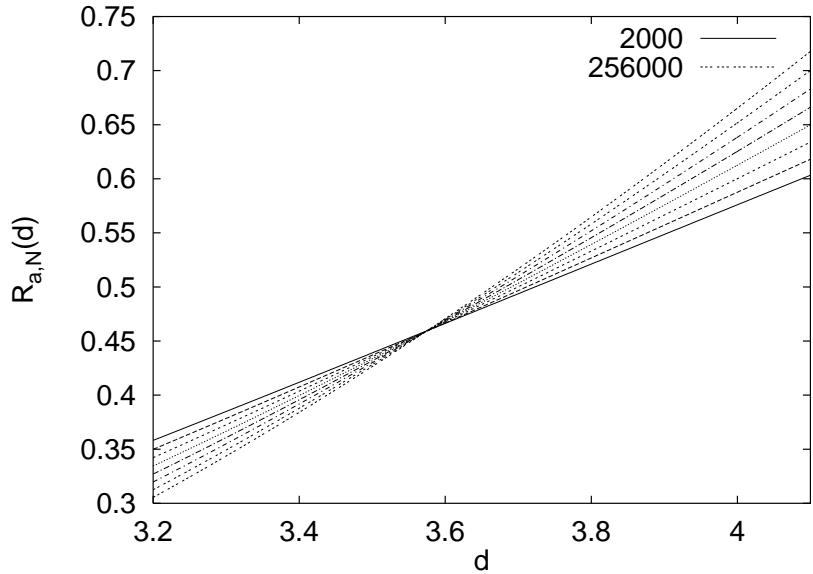,height=8cm,angle=0}}}
\caption[fig1]{The functions $R_{a,N}(d)$ for $N= 2K, 4K, 8K, \ldots, 256K$ 
and for the optimal $a = 0.130$ determined by minimizing Eq.\ \rf{*18}.} 
\label{fig1}
\end{figure}
\begin{figure}
\centerline{\hbox{\psfig{figure=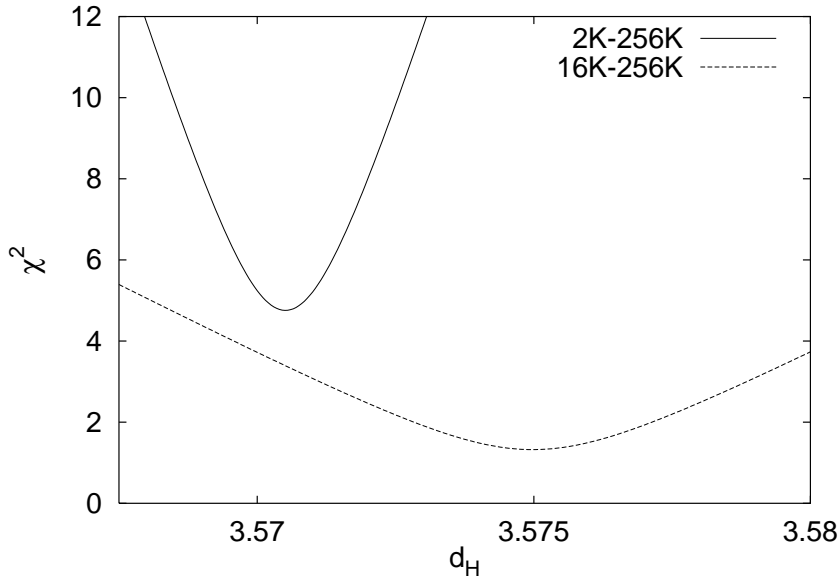,height=8cm,angle=0}}}
\caption[fig2]{$\chi^2(a)$, defined by Eq.\ \rf{*18} for two sets of $N_i$'s:
$N_i = 2K, 4K, \ldots, 256K$ (solid curve) 
and $N_i =16K, 32K, \ldots, 256K$ (dashed curve).}
\label{fig2}
\end{figure}

We have measured the fractal dimension in a number of independent
ways\footnote{The details of these measurements will be reported elsewhere
\cite{us}.}, and various measurements agree. Here we limit ourself 
to report on one particular method, based on the distribution $n_N(r)$.
{}From \rf{*5} we have
\beq{*13}
\la r \ra_N \equiv \frac{1}{N}\sum_{r=0}^\infty r \; n_N(r)  
\sim  N^{1/d_H}\, .
\eeq
Obviously, \rf{*13} could itself serve as a natural definition of $d_H$. 
By measuring $n_N(r)$ we can record $\la r \ra_N$
as a function of $N$ and hence determine $d_H$. 
Strictly speaking, 
we expect this relation to be valid in the limit of infinite $N$, 
while finite $N$ effects will be present in \rf{*13}. 
The finite $N$ effects should be dictated by 
the ``the number of points'' $L$ corresponding 
to the linear size of the system, i.e.\ we expect 
\beq{*14}
\frac{\la r \ra_N}{N^{1/d_H}} 
= {\rm const.} + \frac{{\rm const.}}{L}+ \frac{{\rm const.}}{L^2} + \cdots\, .
\eeq
If we use the fact that $N^{1/d_H}$ {\it is} a typical measure 
for the linear extension of the manifold, 
Eq.\ \rf{*14} can be written, 
\beq{*15}
\la r + a \ra_N \sim N^{1/d_H} + O(\frac{1}{N^{1/d_H}})\, ,
\eeq
by identifying $L = N^{1/d_H}$. 
The parameter $a$, which is considered {\it the shift in} $r$, 
incorporates the next order correction. 
We will discuss possible physical interpretations of the shift $a$ in
detail in \cite{us}. Now, let us define  
\beq{*15a}
R_{a,N}(d) = \frac{\la r + a \ra_{N}}{N^{1/d}}\, .
\eeq
We determine the value 
of $a$ and $d_H$ in the following way: first we measure $\la r \ra_{N_i}$
for a certain number of different volumes $N_i$ of the universes, 
ranging from $N=2K$ to $N=256K$. 
For a given $a$ we choose, for {\it each couple} $N_i,N_j$ of $N$'s, 
the $d_H^{ij}$ such that 
\beq{*16}
R_{a,N_i}(d_H^{ij}) = R_{a,N_j}(d_H^{ij}) \, .
\eeq   
For this choice of $N_i,N_j$ we bin the data and estimate an error 
$\del d_H^{ij}$. Then we determine the average 
\beq{*17}
\bar{d}_H  =  \frac{1}{{\rm \#~pairs}} \sum_{i\neq j} d_H^{ij} \, ,
\eeq
and compute 
\beq{*18}
\chi^2(a) = 
\sum_{i\neq j} \frac{(d_H^{ij} - \bar{d}_H)^2}{(\del d_H^{ij})^2}\, .
\eeq
The preferred pair $(a,d_H(a))$ is determined by the minimum of $\chi^2(a)$.
This method works  quite impressively.  
In Fig.\ \ref{fig1} we have shown the intersection of the curves 
$R_{a,N}(d)$ as a function of $d$ for the optimal choice of $a$. 
%
%
The important point is that there exists a value of $a$ 
where the curves intersect with high precision and that the   
the range of $a$ where $\chi^2(a)$ is acceptably small,
i.e.\  $O(1)$, is quite small and hence $d_H$ will be determined with 
high precision. 
In Fig.\ \ref{fig2} we show $\chi^2(a)$.  
In this way we get 
\beq{*19}
a_m= 0.139\pm 0.005\, ,~~~~~d_H(a_m) = 3.574\pm 0.003\, .
\eeq
In \rf{*19} we have estimated the error as follows. 
Define an interval of acceptance $[a_{\rm min},a_{\rm max}]$ of $a$ 
by demanding that $\chi^2(a) < 2 \chi^*$ 
where $\chi^* = {\rm max}\{1 , \chi^2(a_m)\}$
and find the variation of $d(a)$ in this interval.
After this we repeat the whole procedure by making various cuts in the 
pairs of $N_i$'s included in \rf{*17} and \rf{*18}, 
discarding successively the smallest $N_i$'s. 
The value of $d_H$ in Eq.~\rf{*19} 
agrees with the original value $d_H = 3.5\pm 0.2$ \cite{kketal}. 
The strength of finite-size scaling is that 
one can obtain higher precision results with 
the use of much smaller lattices. 
The original simulation needed $5000K$ size lattices.

\subsection{The boundary}
   
\begin{figure}
\centerline{
\epsfxsize=3.0in \epsfysize=2in \epsfbox{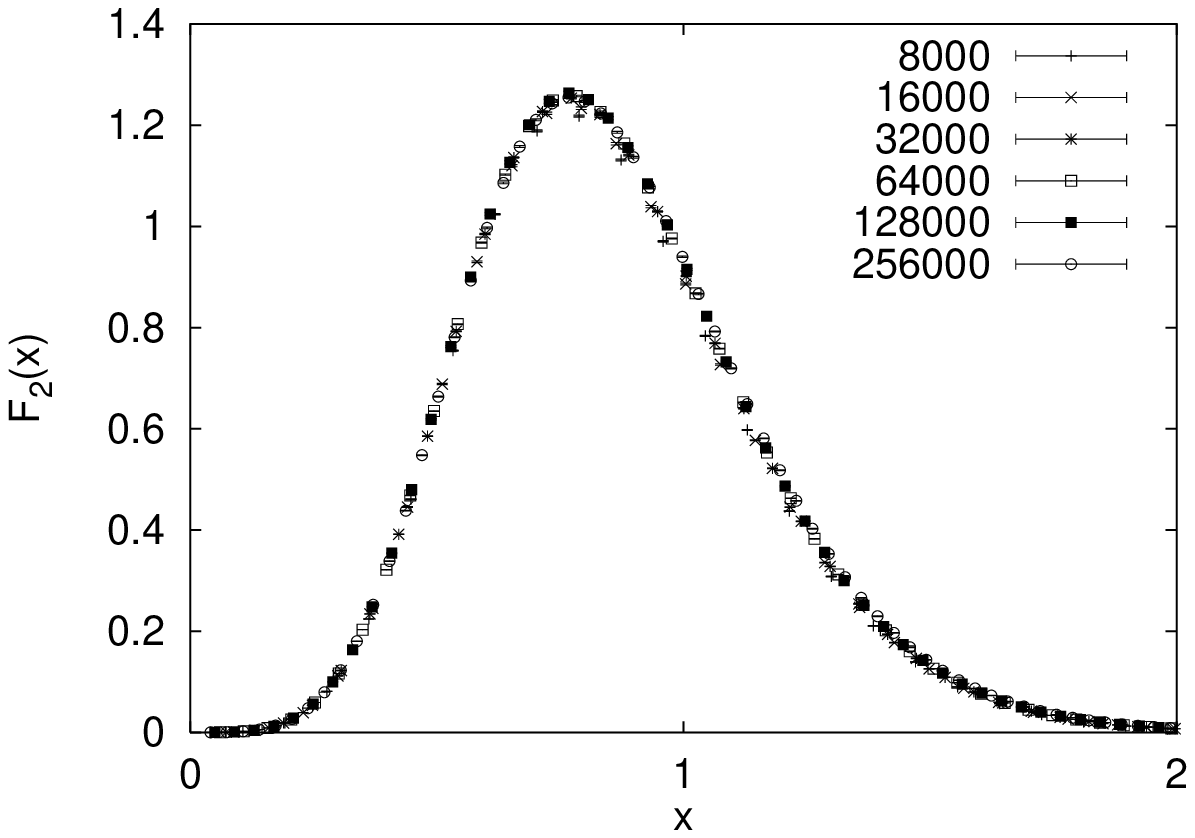}
\epsfxsize=3.0in \epsfysize=2in \epsfbox{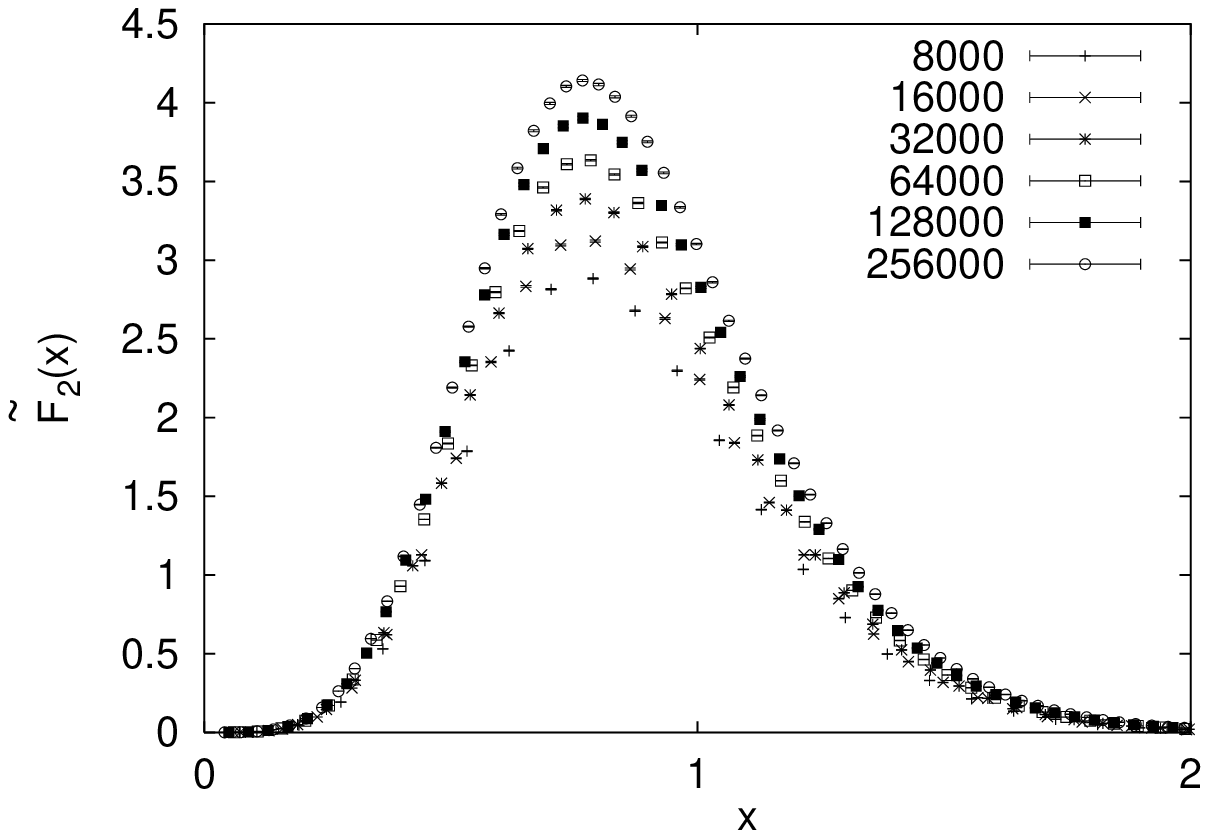}}
\caption[fig3]{The scaling functions $F_n(x)$ defined by Eq.\ \rf{*21}
(left figure) and the scaling functions $\tilde{F}_n(x)$ defined by Eq.\
\rf{*20}, for $n=2$ and $N= 8K, 16K, \ldots, 256K$.} 
\label{fig3}
\end{figure}

We now turn to the measurements of $\la l^n \ra_{r,N}$. 
These observables are constructed from $\rho_N (l,r)$, which 
can readily be measured in the simulations. 
If ${\rm dim}[ \, N \, ]={\rm dim}[ \, l^2 \, ]$, 
then from scaling arguments, we expect
\beq{*20}
\la l^n \ra_{r,N}  \sim  N^{n/2} \tilde{F}_n (x)\, ,
~~~~~x = \frac{r}{N^{1/d_H}}\, .
\eeq
However, our measurements are consistent with 
the following scaling relation 
\beq{*21}
\la l^n \ra_{r,N}  \sim  N^{2n/d_H} F_n(x),
~~~~~{\rm for}~~n \geq 2\, , 
\eeq
which implies that ${\rm dim}[ \, l \, ]={\rm dim}[ \, r^2 \, ]$. 
Eq.\ \rf{*21} indicates that we have 
\beq{*22}
\la l^n \ra_{r,N} \sim r^{2n}
~~~~~{\rm for}~~1 \ll r \ll N^{1/d_H},~n \geq 2\, .
\eeq
Again, relations like \rf{*21} are expected to be valid up to finite size 
effects, as in Eq.\ \rf{*14}. 
As a first phenomenological correction 
we use a shift $r \to r+a$ as in \rf{*15} to find the best scaling 
function $F_n(x)$ for a suitable range of $N_i$'s. 
In Fig.\ \ref{fig3}
we have shown $F_n(x)$ for $n=2$ for the values of $a$ which provide
the best scaling function. 
Similar pictures exist for $n=3$ and $n=4$.  
This is to be compared to the scaling given in Eq.\ \rf{*20}.  
We see that it is not possible to find a scaling function
$\tilde{F}_n(x)$ if we used the {\it ansatz} \rf{*20}.  
Assuming the scaling \rf{*21} 
we get an independent determination of $d_H = 3.63\pm0.04$. 
This result is remarkably consistent with the value of $d_H$ 
determined by the other methods we used, 
considering the systematic errors due to finite size effects. 
One, however, should consider the possibility of a dimensional relation 
of the form: 
\beq{*22a} 
{\rm dim}[ \, l \, ]={\rm dim}[ \, r^{2(1-\epsilon)} \, ] \, .
\eeq
In this case, for given $d_H$, one can use the relation 
$\la l^n \ra_{r,N} = N^{2n(1-\epsilon)/d_H}F_n(x)$ 
in order to determine the value of $\epsilon$.  
Using the lowest value we obtained for $d_H$ using other
methods, we obtain $\epsilon < 0.03$. 
Details will be published elsewhere \cite{us}.

\subsection{The spectral dimension}

\begin{figure}
\centerline{\hbox{\psfig{figure=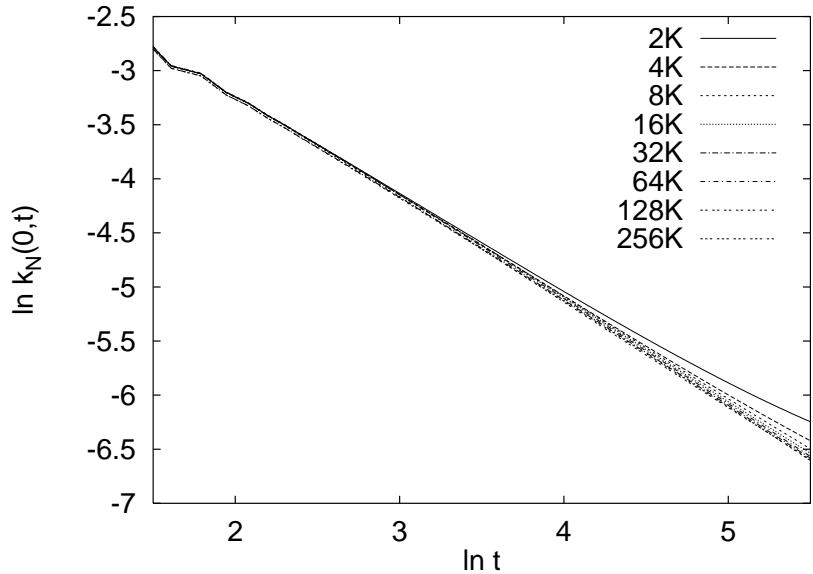,height=8cm}}}
\caption[fig4]{A plot of $ \ln k_N(0,t)$ versus $\ln t$ 
for $N = 2K, 4K, \ldots, 256K$.} 
\label{fig4}
\end{figure}

Let us finally turn to the spectral dimension $d_s$, as defined by \rf{*10}.
For a given triangulation $k_N(0,t)$ can be calculated by means 
of \rf{*6} and \rf{*7}, and the quantum average is then obtained 
by performing the average over different triangulations, since these
are generated with the correct weight by the recursive sampling.
In Fig.\ \ref{fig4} 
we have shown a plot of $\ln k_N(0,t)$ versus $\ln t$. 
When $t$ is too small, discretization effects interfere with the form \rf{*10}, 
as discussed in details in \cite{ajw}.  
For larger values of $t$ it is possible to perform a very good fit to \rf{*10}, 
as is clear from Fig.\ \ref{fig4}. 
For the $256K$ lattice the value of $d_s$ is
\beq{*23} 
d_s = 1.980 \pm 0.014\, .  
\eeq 
Due to finite size effects, the value of $d_s$ depends on the lattice
size, but it clearly approaches $2$ as $N$ becomes large. Exploiting the
$N$ dependence of $k_N(r,t)$ it is possible to determine the exponents
$\l$ and $\sg$ introduced in Eqs.\ \rf{*10} and \rf{*11}. We find good
agreement with the scaling prediction \rf{*smoothcond2}. A detailed analysis of
the diffusion equation and the numerical verification of \rf{*smoothcond2} will
be published elsewhere.

\section{Discussion}

The fractal structure of quantum gravity coupled to matter
has provided us with somewhat of a puzzle. 
The first analytic result suggested that \cite{dhk}
\beq{*24}
d_H = - \frac{2}{\g_{\rm str}} \, , 
~~~~~
\g_{\rm str} = \frac{1}{12} \Bigl( \, c - 1 - \sqrt{(25-c)(1-c)} \, \Bigr) \, .
\eeq
This formula has later been obtained in a number of different ways 
using the string-field approach developed in \cite{kawai.et.al}. 
There are several problems with the ``proofs''. 
For instance, the string-field proof is based on 
the identification of proper time with geodesic distance. 
However, only in the case of pure two-dimensional gravity 
one can clearly identify the proper time used in 
string-field theory with the geodesic distance. 
Further, the formula \rf{*24} predicts that 
$d_H \to \infty$ for $c\to 1$ and 
$d_H \to 0$ for $c\to -\infty$. 
The latter result is clearly undesirable 
since one expects that  $d_H \to 2$ in the semi-classical limit 
$c \to -\infty$. 
In fact \rf{*24} predicts $d_H=2$ in the case considered here, 
$c=-2$ ($\g_{\rm str} = -1$), 
and a very recent constructive proof,
using an explicitly constructed transfer matrix for $c=-2$,  
confirms this prediction \cite{akw}. 
The prediction is clearly in disagreement with the computer measurements
reported in this article. In a similar way the prediction \rf{*24} 
for $c>0$ has been in contradiction with the Monte Carlo simulations 
performed for the Ising model ($c=1/2$, $\g_{\rm str} = -1/3$) and the 
three-states Potts model ($c=4/5$, $\g_{\rm str}=-1/5$). In these cases one could 
argue that since the fractal dimension predicted is so large, the 
systems used in the computer simulations have been too small to 
observe the correct fractal dimensions. Although it is hard to 
understand how one can get all predicted {\it KPZ}-exponents correct
in the numerical simulations, 
but not being able to measure the fractal dimension, one could not 
rule out entirely this criticism. The important point is that
the criticism is not valid for the present numerical investigation,
since the predicted $d_H$ is small (namely 2), 
and we are able to deal with very large systems. 

The alternative prediction \cite{ksw} for $d_H$ is 
\beq{*25}
d_H  =  -2\frac{\a_1}{\a_{-1}} = 
2 \times \frac{\sqrt{25-c}+\sqrt{49-c}}{\sqrt{25-c}+ \sqrt{1-c}}\, .
\eeq
The origin of this equation is to be found 
in the analysis of the diffusion equation in Liouville theory \cite{ksw}
and is based on the observation that \rf{*8}-\rf{*smoothcond2} imply that 
\beq{*26}
{\rm dim}[ \, \la r^2(t) \ra_N \, ] = {\rm dim}[ \, N^{2/d_H} \, ] \, .
\eeq
This follows from the assumed scaling, 
independent of the specific model of dynamical triangulations. 
On the other hand, in Liouville theory 
one can use the De-Witt short distance expansion of 
the heat kernel in terms of geodesic distance to deduce \cite{ksw} that
\beq{*27}
{\rm dim}[ \, \la r^2(t) \ra_N \, ] = {\rm dim}[ \, N^{-(\a_{-1}/\a_1)} \, ] \, ,
\eeq 
provided $r$, $N$ and $t$ are viewed as continuum geodesic distance, 
continuum area and continuum diffusion time in Liouville theory. 
In \rf{*27} $\a_{-n}$ denotes the gravitational dressing of 
a $(n+1,n+1)$ conformal field, i.e.\
$$
\int\! d^2\xi \sqrt{g}\Phi_{n+1}(g) \to
\int\! d^2\xi \sqrt{\hat{g}} \; e^{\a_{-n} \phi} \Phi_{n+1}(\hat{g}) \, , 
~~~~{\rm for}~~
g_{\mu\nu}(\xi) = e^{\phi(\xi)} \hat{g}_{\mu\nu}(\xi) \, ,
$$
where $\hat{g}_{\mu\nu}(\xi)$ is the background metric. 
The requirement that 
$e^{\a_{-n}\phi}\Phi_{n+1}(\hat{g})$ is a $(1,1)$ conformal field fixes 
\beq{*28}
\a_n = \frac{2n}{1+\sqrt{(25-c-24n)/(25-c)}}\, .
\eeq
For $c=0$ one obtains $d_H = 4$ 
(in agreement with the transfer matrix prediction), 
while for $c=-2$ the prediction is 
\beq{*29}
  d_H(c=-2)  =  (3+\sqrt{17})/2 = 3.561\ldots \, .
\eeq
It agrees with the numerical results reported in this paper.

However, clearly this is not the complete story 
since we have also obtained 
\beq{*30}
{\rm dim}[ \, l^n \, ] = {\rm dim}[ \, r^{2n} \, ] \, ,
~~~~~{\rm for}~~n>1 \, .
\eeq
The same result is valid for $c=0$ and from numerical simulations
for $0< c< 1$ it  seems to be valid in this region, too \cite{aa}.
The reason such a result can appear is apparent from the transfer
matrix calculations for $c=0$. 
In this case we have
\beq{*31}
\rho_N(r,l) 
= \frac{c_1}{r^2} G(l/r^2) +
  c_2 \ep^{-3/2} r^{3}(2-l/r^2)\del(l-\sqrt{3}\ep) \, , 
\eeq
where $\ep$ is a cut-off (the lattice spacing) and 
\beq{*32}
G(z)= \Bigl(z^{-5/2} + \oh z^{-3/2} + 
       \frac{14}{3} z^{1/2} \Bigr) \; e^{-z} \, ,
~~~~~{\rm for}~~N \to \infty \, .
\eeq
For low moments, $n=0$ and 1, the terms  in \rf{*31} which are singular 
for $l \to 0$ will dominate the evaluation of 
\beq{*33}
\la l^n \ra_{r,N} = \int_\ep dl \; l^n \, \rho_N(r,l) \, ,
\eeq
while for $n \geq 2$ these terms become integrable 
and the cut-off dependence vanishes. 
It seems that the situation is the same at least for $c \in [-2,1)$.
The exists a regular part, $\rho_N^{({\rm reg})}(r,l)$, 
of $\rho_N(r,l)$ such that 
\beq{*34}
d l \rho_N^{({\rm reg})} (r,l) = d (l/r^2) G(l/r^2)
\eeq
is a function only of $l/r^2$, and which dominates the integral \rf{*33} 
for $n \geq 2$, while a part, singular for $l \to 0$, 
dominates \rf{*33} for $n=0$ and $1$. 

In order to fully {\it understand} the concept of fractal dimension 
we still have to provide an explanation of an expression like 
\rf{*31} for $c \neq 0$.

\subsection*{Acknowledgments}
J.A. acknowledges the support of the Professor Visitante Iberdrola
grant and the hospitality at the University of Barcelona, where part
of this work was done.

\end{document}